\documentstyle{rnote}
\author{L.~Didukh}
\address{Ternopil State Technical University, 
Department of Physics\\ 
56 Rus'ka Str., Ternopil UA--46001, Ukraine \\
E-mail: didukh@tu.edu.te.ua}

\title{A new two-pole approximation in the Hubbard model. Metal-insulator 
transition}

\submitted{December 23, 1999}

\begin{document}

\maketitle

Two-pole approaches in the Hubbard model~\cite{hubb1} and the Hubbard bands 
conception (being the consequence of a two-pole approximation) have been 
useful for understanding of the peculiarities of electric and magnetic
properties of narrow-band materials~\cite{hubb1}-\cite{did}. However 
within the framework of two-pole approaches there are series of issues,
in particular the problem of metal-insulator transition description.
 
In this paper a new two-pole approximation, which allows to describe
the transition from an insulating state to a metallic one at increase of
bandwidth, and also the observable in some compounds transition from a
metalic state to an insulating one with increasing temperature, is 
presented.

Let us take the operators of creation and destruction
of an electron with spin $\sigma$ ($\sigma=\downarrow,\uparrow$) on 
$i$-site in the form:
\begin{eqnarray}
a_{i\sigma}^{+}=d_{i\sigma}^{+}+h_{i\sigma}, \quad
a_{i\sigma}=d_{i\sigma}+h_{i\sigma}^{+},
\end{eqnarray}
where 
$d_{i\sigma}^{+}=a_{i\sigma}^{+}n_{i\bar{\sigma}},\
d_{i\sigma}=a_{i\sigma}n_{i\bar{\sigma}},\
h_{i\sigma}^{+}=a_{i\sigma}(1-n_{i\bar{\sigma}}),\
h_{i\sigma}=a_{i\sigma}^{+}(1-n_{i\bar{\sigma}}),$
and $\bar{\sigma}$ denotes the projection of electron spin opposite to
$\sigma$. Between the $d$-$h$-operators and Hubbard 
$X$-operators~\cite{hubb} direct relations exist~\cite{ds}.
The Hubbard Hamiltonian~\cite{hubb1} in terms of the $d$-$h$-operators is 
written as
\begin{eqnarray}
&&H= H_{0}+H_{1}+H'_{1},
\\
&&H_{0}=-{\mu \over 2}\sum_{i\sigma}\left(d_{i\sigma}^{+}d_{i\sigma}+
h_{i\sigma}h_{i\sigma}^{+}\right)
+{U \over 2}\sum_{i\sigma}d_{i\sigma}^{+}d_{i\sigma}, \\
\label{H1}
&&H_{1}=t\sum \limits_{ij\sigma,i\neq j}\left(d_{i\sigma}^{+}d_{j\sigma}-
h_{i\sigma}^{+}h_{j\sigma}\right),
\quad
H_{1}'= t\sum \limits_{ij\sigma,i\neq j}\left(d_{i\sigma}^{+}h_{j\sigma}^{+}+
h.c.\right),
\end{eqnarray}
where $\mu$ is the chemical potential; 
$U$ is the intra-atomic Coulomb repulsion;
$t$ is the nearest-neighbor hopping integral.
$H_{0}$ describes system in the atomic limit,
$H_{1}$ describes electron hoppings between doubly occupied sites
(with two electrons of opposite spins - doublons) and single 
occupied sites (the first sum in $H_1$)
and electron hoppings between single occupied sites and empty 
sites (holes) (the second sum in $H_1$).
$H'_{1}$ describes ``hybridization'' between the ``h-band'' and ``d-band''
(the processes of pair creation and annihilation of holes and doublons).
The structure of the Hamiltonian originates the approximation given below.

The Green functions $\langle\langle d_{p\uparrow}|
d_{s\uparrow}^{+}
\rangle\rangle$ and $\langle\langle h_{p\uparrow}^{+}|d_{s\uparrow}^{+}
\rangle\rangle$ satisfy the equations:
\begin{eqnarray}
&&(E+\mu-U)\langle\langle d_{p\uparrow}|d_{s\uparrow}^{+}\rangle\rangle=
{n_{\uparrow}\over 2\pi}\delta_{ps}+
\langle\langle [d_{p\uparrow}, H_1]_{-}|d_{s\uparrow}^{+}\rangle\rangle
\nonumber\\
&&+t\sum_i\langle\langle (h_{p\downarrow}h_{p\uparrow}^{+}h_{i\downarrow}^{+}
-h_{p\uparrow}^{+}d_{p\downarrow}d_{i\downarrow}^{+}-
n_{p\downarrow}h_{i\uparrow}^{+})|d_{s\uparrow}^{+}\rangle\rangle ,
\\
&&(E+\mu)\langle\langle h_{p\uparrow}^{+}|d_{s\uparrow}^{+}\rangle\rangle=
\langle\langle [h_{p\uparrow}^{+}, H_1]_{-}|d_{s\uparrow}^{+}\rangle\rangle
\nonumber\\
&&+t\sum_i\langle\langle (h_{p\downarrow}h_{p\uparrow}^{+}d_{i\downarrow}
-h_{p\uparrow}^{+}d_{p\downarrow}h_{i\downarrow}
-(1-n_{p\downarrow})d_{i\uparrow})|d_{s\uparrow}^{+}\rangle\rangle ,
\end{eqnarray}
with $n_{p\sigma}=a_{p\sigma}^{+}a_{p\sigma}$, $n_{\sigma}=
\langle n_{p\sigma}\rangle$ and $[A, B]_{-}=AB-BA$. To
obtain the closed system of equations for the Green functions 
$\langle\langle d_{p\uparrow}|d_{s\uparrow}^{+}
\rangle\rangle$ and $\langle\langle h_{p\uparrow}^{+}|d_{s\uparrow}^{+}
\rangle\rangle$ we suppose
\begin{eqnarray}
\langle\langle [d_{p\uparrow}, H_1]_{-}|d_{s\uparrow}^{+}\rangle\rangle=
\sum_j\varepsilon(pj)\langle\langle d_{j\uparrow}|d_{s\uparrow}^{+}
\rangle\rangle, 
\\
\langle\langle [h_{p\uparrow}^{+}, H_1]_{-}|d_{s\uparrow}^{+}\rangle\rangle=
\sum_j\varepsilon_1(pj)\langle\langle h_{j\uparrow}|d_{s\uparrow}^{+}
\rangle\rangle,
\end{eqnarray} 
where $\varepsilon(pj),\ \varepsilon_1(pj)$ are the non-operator 
expressions which are calculated by the method decribed in Ref.~\cite{did}.
At electron concentration $n=1$ (this is the important situation to study
metal-insulator transition) in a paramagnetic state 
($n_{\uparrow}= n_{\downarrow}$) we have
$\varepsilon(pj)=\varepsilon_1(pj)=(1-2c)t$, with $c$ being 
the concentration of polar states (holes or doublons). 
For the model Hamiltonian $H_0+H_1$ this approximation leads to the 
criterion of metal-insulator transition $(U/2w)_{cr}=1$ reproducing
the exact result of Ref.~\cite{ovch}.

Among the Green function originated from the expressions 
$\langle\langle [d_{p\uparrow}, H_{1}']_{-}|d_{s\uparrow}^{+}\rangle\rangle$
and 
$\langle\langle [h_{p\uparrow}^{+}, H_{1}']_{-}|d_{s\uparrow}^{+}\rangle\rangle$
we take into account the ``diagonal'' Green function only. 
The decoupling procedure of these Green functions is made by means of 
the mean-field approximation:
\begin{eqnarray}  
\langle\langle n_{p\downarrow}h_{i\uparrow}^{+}|
d_{s\uparrow}^{+}\rangle\rangle\simeq
1/2\langle\langle h_{i\uparrow}^{+}|d_{s\uparrow}^{+}\rangle\rangle,
\qquad
\langle\langle (1-n_{p\downarrow})d_{i\uparrow})|d_{s\uparrow}^{+}
\rangle\rangle\simeq 1/2\langle\langle d_{i\uparrow}^{+}|
d_{s\uparrow}^{+}\rangle\rangle ;
\end{eqnarray}
by making these approximations we neglect the processes describing 
the ``inter-band'' hoppings of electrons which are connected with
spin turning over and ``inter-band'' hoppings with 
creation or annihilation of two electrons on the same site.

Finally, in {\bf k}-representation one-electron Green function is
\begin{eqnarray} \label{gf}
&&\langle\langle a_{p\uparrow}|a^{+}_{s\uparrow}\rangle\rangle_{\bf k}=
{1\over 4\pi}\left({A_{\bf k}\over E-
E_1({\bf k})}+{B_{\bf k}\over E-E_2({\bf k)}}\right),\\
&&A_{\bf k}=1-{t({\bf k})\over \sqrt{U^2+t^2({\bf k})}}, \quad
B_{\bf k}=1+{t({\bf k})\over \sqrt{U^2+t^2({\bf k})}}, \\
&&E_{1,2}({\bf k})=(1-2c)t({\bf k})\mp {1\over 2}
{\sqrt{U^2+t^2({\bf k})}} \label{sp},
\end{eqnarray}
where $t({\bf k})$ is the hopping integral in ${\bf k}$-representation;
$E_1({\bf k})$ is the quasiparticle energy spectrum in the lower Hubbard 
band ($h$-band), $E_2({\bf k})$ is the quasiparticle energy spectrum 
in the upper Hubbard band ($d$-band).
One-electron Green function~(\ref{gf}) and energy spectrum~(\ref{sp})
are exact in the atomic and band limits. 
 
The peculiarity of energy spectrum~(\ref{sp}) is the dependence on polar 
states concentration (and on temperature), that differs it from
the two-pole approximations of Hubbard~\cite{hubb1}, and Ikeda, Larsen, 
Mattuck~\cite{iked}. A distinction of the proposed approximation from 
the approximations based on the ideology of Roth~\cite{roth} 
(in this connection 
see also Refs.~\cite{herr}-\cite{izyu}) is, first of all, 
ability of the proposed 
approximation to describe metal-insulator transition. Spectrum~(\ref{sp}) 
differs also from the spectrum earlier obtained by the author~\cite{did} by
presence of term $\sqrt{U^2+t^2({\bf k})}$ instead of the expression
$\sqrt{U^2+4c^2t^2({\bf k})}$ obtained in Ref.~\cite{did}. This leads to
series of the distinctions of results of this work from the results of 
work~\cite{did} ($c(U/w)$--dependence, the condition of metal-insulator 
transition); for doped Mott-Hubbard materials, when electron concentration 
$n \neq 1$, and we can restrict ourselves to consideration one of
two Hubbard's bands, approach of Ref.~\cite{did} and proposed in 
this work approach are equivalent.

The concentration of polar states is calculated with the help of the 
Green function 
\begin{eqnarray}
&&\langle\langle d_{p\uparrow}|d_{s\uparrow}^{+}\rangle\rangle_{\bf k}=
{1\over 8\pi}\left({C_{\bf k}\over E-
E_1({\bf k})}+{D_{\bf k}\over E-E_2({\bf k)}}\right),\\
&&C_{\bf k}=1-{U\over \sqrt{U^2+t^2({\bf k})}}, \quad
D_{\bf k}=1+{U\over \sqrt{U^2+t^2({\bf k})}}.
\end{eqnarray}
At $T=0$ we have
\begin{eqnarray}
c={1\over 4}+{U\over 8w}\ln{1-4c\over 3-4c}
\end{eqnarray}
if $U/2w\leq (U/2w)_{cr}$ with $(U/2w)_{cr}=0.836$
($w=z|t|$, $z$ is the number of nearest neighbors to a site); 
for $U/2w>(U/2w)_{cr}$ 
the concentration of polar states is
\begin{eqnarray}
c={1\over 4}+{U\over 8w}\ln{-w+\sqrt{U^2+w^2}\over w+\sqrt{U^2+w^2}}.
\end{eqnarray}

Expression~(\ref{sp}) describes the vanishing of energy gap in the
spectrum of paramagnetic insulator at increasing $w$ (under pressure).
Really, the energy gap width (difference  of  energies  between 
bottom of the upper and top of the lower Hubbard bands)
\begin{eqnarray} \label{eg}
\Delta E=-2(1-2c)w+\sqrt{U^2+w^2}.
\end{eqnarray}
vanishes when the condition  $(U/2w)_{cr}=0.836$ is satisfied. This 
value is close to the result of ``Hubbard-III'' approximation~\cite{hubb3}.
It is important to note that in the point of gap disappearance we have
$c\neq 0$.

Energy gap~(\ref{eg}) is 
temperature dependent. This allows to explain
observed in some narrow-band materials transition from a metallic 
to an insulating state with increasing temperature (see, for 
example~\cite{mott}).

Above we have considered the case of a paramagnetic narrow-band material at
half-filling ($n=1$). It is interesting to study non-halffilled case 
($n\neq 1$); here the energy spectrum can be essentially modified by the
spin-dependent (in general case) shifts of the Hubbard band centers (on the
analogy with the Harris and Lange result~\cite{harr}). This case 
(which is important
to study ferromagnetism in narrow energy bands) will be considered
in subsequent paper (as well as the problem of antiferromagnetism).

\end{document}